\documentclass[12pt]{article}
\usepackage{amsmath,amssymb}
\usepackage{graphicx}
\usepackage[english]{babel}
\usepackage{cite}

\newcommand{\nc}{\newcommand}

\newcommand{\be}{\begin{equation}}
\newcommand{\ee}{\end{equation}}
\newcommand{\bes}{\begin{equation*}}
\newcommand{\ees}{\end{equation*}}

\newcommand{\bea}{\begin{eqnarray}\displaystyle}
\newcommand{\eea}{\end{eqnarray}}

\nc{\NN}{{\cal N}}
\nc{\OO}{{\cal O}}
\nc{\KK}{{\cal K}}
\nc{\PP}{{\cal P}}
\nc{\UU}{{\cal U}}
\nc{\TO}{{\cal T}}
\nc{\KB}{{\cal K}}
\nc{\AN}{{\cal A}}
\nc{\MS}{{\cal M}}
\nc{\CC}{{\cal C}}
\nc{\DD}{{\cal D}}
\nc{\HH}{{\cal H}}
\nc{\GG}{{\cal G}}
\nc{\YY}{{\cal Y}}
\nc{\VV}{{\cal V}}
\nc{\II}{{\cal I}}

\nc{\BZ}{{\cal Z}}

\nc{\CY}{C{\rm{Y}_3}}
\nc{\ox}{\otimes}
\nc{\x}{\times}
\nc{\w}{\wedge}
\nc{\W}{\bigwedge}
\nc{\p}{\partial}
\nc{\bp}{\bar{\partial}}
\nc{\pbar}{\bar{\partial}}
\nc{\MM}{{\sf M}}
\nc{\wt}{\widetilde}
\nc{\wh}{\widehat}
\nc{\vp}{\varphi}
\nc{\ep}{\epsilon}
\nc{\vep}{\varepsilon}
\nc{\vtheta}{\vartheta}

\nc{\RR}{\mathbf{R}}
\nc{\RE}{{\cal R}}
\nc{\Z}{\mathbf{Z}}
\nc{\ZZ}{\mathbf{Z_2}}
\nc{\RP}{\mathbb{R}\mathbb{P}}
\nc{\cplx}{\mathbf{C}} 
\nc{\one}{{\mathbf{1}}}

\numberwithin{equation}{section}

\usepackage{babel}
\makeatother

\begin{document}

\vskip 1cm \centerline{\large {\bf On triviality of $\lambda\phi^{4}$ quantum field theory in four
dimensions}}
\vskip 1cm
\renewcommand{\thefootnote}{\fnsymbol{footnote}}
\centerline{{\bf Dmitry I. Podolsky\footnote{podolsky@phys.cwru.edu}}}

\vskip .5cm \centerline{\it ${}^{1}$Department of Physics \& CERCA, Case Western Reserve University } \centerline{\it
10900 Euclid Ave., Cleveland, OH 44106} 

\setcounter{footnote}{0}
\renewcommand{\thefootnote}{\arabic{footnote}}

\begin{abstract}
Interacting quantum scalar field theories in $dS_{D}\times M_{d}$
spacetime can be reduced to Euclidean field theories in $M_{d}$ space
in the vicinity of $I_{+}$ infinity of $dS_{D}$ spacetime. Using
this non-perturbative mapping, we analyze the critical behavior of Euclidean
$\lambda\phi_{4}^{4}$ theory in the symmetric phase and find the
asymptotic behavior $\beta(\lambda)\sim\lambda$ of the beta function
at strong coupling. Scaling violating contributions to the beta function
are also estimated in this regime.
\end{abstract}

\newpage

\section{Introduction}

Almost 35 years ago\footnote{Actually, the history of triviality proposal begins somewhat earlier than that: Lev Landau and his students were discussing zero charge behavior of pseudoscalar theories already in 1955 \cite{Landau}.} Wilson and Kogut \cite{WilsonKogut} have suggested
that the $\lambda\phi_{d}^{4}$ quantum field theory is in fact trivial
in number of dimensions $d\ge4$, \emph{e.g.}, the renormalized coupling
$\lambda(p)$ vanishes at small momenta $p$ for any finite bare
coupling $\lambda_{0}$ and the correlation properties of the field
$\phi$ become Gaussian. The question whether the theory is trivial
in this sense is not merely academic: one of the strongest theoretical
upper bounds on the value of the Higgs mass $m_{H}$ comes from the fact
that the quantum field theory of the Higgs field coupled to non-Abelian
gauge theory describing weak interaction is either trivial or asymptotically
free depending on the value of $m_{H}$ \cite{HiggsMass}.

Triviality of the theory in $d\ge5$ dimensions was rigorously established
\cite{Aizenman1,Aizenman2} using the representation of Euclidean field
theories as systems of random currents (essentially world lines of
elementary excitations in the first quantized version of the theory),
but the case $d=4$ turned out to be special and much more complicated
\cite{TrivialityArgumentsRG,TrivialityArgumentsMath}. Although the
validity of the Wilson conjecture in this case is supported by results
of lattice simulations \cite{TrivialityNumerics}, to this day there
exists no strict proof of triviality of the $\lambda\phi_{4}^{4}$
theory.

Let us recall why the $d=4$ case is so special in both the field theoretic
framework and the geometric picture of random currents. In the latter,
triviality is equivalent to the statement that two random Brownian
paths generically do not intersect. As it turns out, the probability
of such an intersection is indeed zero if Brownian paths are located in space
with number of dimensions higher than $4$. In $d=4$ dimensions, the geometry
of a random walk is more complicated: although the probability for
two Brownian paths to intersect is zero, the probability for both
of them to penetrate a ball of the radius $\epsilon$ is $1$ for
any, even infinitely small, value of $\epsilon$.

In the field theoretic framework, the triviality of $\lambda\phi_{4}^{4}$
theory is directly related to the existence of a non-perturbative Landau
pole. The 1-loop renormalization law 
\be
\lambda(p')=\frac{\lambda(p)}{1-\frac{3\lambda^{2}(p)}{16\pi^{2}}\log\frac{p'}{p}},
\label{eq:LandauPole}
\ee
featuring a singularity at finite momentum $p'\sim p\cdot\exp\left(\frac{16\pi^{2}}{3\lambda^{2}(p)}\right)$,
implies that the renormalized coupling vanishes at small momenta for
any finite value of the bare coupling $\lambda_{0}$. On the other
hand, since one expects the coupling $\lambda$ to grow in the vicinity
of the Landau pole, the validity of the 1-loop approximation for the
beta function
\be
\beta(\lambda)=\frac{3}{16\pi^{2}}\lambda^{2}
\ee
breaks down. It is not completely clear whether the Landau pole survives
at the non-perturbative level. The alternative to the runaway behavior
(\ref{eq:LandauPole}) would be the existence of a non-perturbative UV fixed
point(s) and, as a consequence, asymptotic freedom at very large momenta.
Results of lattice simulations seem to disfavor such a scenario \cite{TrivialityNumerics}.

In the absence of UV fixed points, one expects that the perturbative
formulation of the $\lambda\phi_{4}^{4}$ theory is UV incomplete
since it is impossible to probe the theory at arbitrarily large momenta.
A question appears in this context whether it is possible to find
an embracing {}``fundamental'' theory which at small momenta reduces to
the strongly coupled $\lambda\phi_{4}^{4}$ theory with $\lambda(p)\gg1$
and negligible scaling violation corrections \cite{TrivialityArgumentsRG}.
In the present work we address this question and present a corresponding
embracing theory explicitly. We find the asymptotic behavior $\beta(\lambda)\sim\lambda$
at large values of the renormalized coupling, implying that the Landau
pole is absent at the non-perturbative level.

The paper is organized as follows. In the Sections \ref{sec:stochastic}
and \ref{sec:UVphysics} we study correlation properties of scalar
quantum field theories in $dS_{D}\times M_{d}$ spacetime, where $M_{d}$
is a $d$-torus or $d$-dimensional non-compact flat Euclidean space.
We find that the Nicolai map is naturally generated for such theories
relating the large scale (superhorizon) scalar field $\Phi$ and the
{}``random force'' composite operator - a superposition of modes
with physical wavelength of the order of the $dS$ horizon size. Using
the Nicolai map we explicitly derive a correspondence between self-interacting
scalar QFTs in $dS_{D}\times M_{d}$ spacetime and \emph{Euclidean}
field theories in $M_{d}$ target space: the former are reduced to
the latter in the vicinity of the $I_{+}$ infinity of de Sitter space.

This is an important byproduct of our study. In a sense, the correspondence
between field theories in $dS_{D}\times M_{d}$ and $M_{d}$ spaces
means that de Sitter space provides a natural geometric framework
for stochastic quantization of a self-interacting Euclidean scalar
field theory in $d$-dimensional space. The procedure of stochastic quantization
first introduced by Parisi and Wu to analyze correlation properties
of Euclidean gauge theories without gauge fixing \cite{ParisiWu},
has led to development of new effective algorithms for lattice QFT
simulations, advances in study of large $N$ theories and master fields,
etc. \cite{DamgaardHuffel} Recently, it was discovered that several
apparently unrelated QFTs such as the Wess-Zumino-Novikov-Witten model
and the strong coupling limit of topologically massive Yang-Mills theory
are actually connected via stochastic quantization \cite{Dijkgraaf}.
It would be interesting to understand in this context whether the
developed geometric framework for stochastic quantization implies
the existence of a field theory/gravity duality relating the field theories
listed in \cite{Dijkgraaf}.

In the Section \ref{sec:Critical}, the derived correspondence is used
to study critical properties of $\lambda\phi_{4}^{4}$ theory.
It allows us to introduce a UV completion of $\lambda\phi_{4}^{4}$
theory such that its correlation properties at large momenta are governed
by a higher dimensional theory. We estimate the beta function of the
theory at strong coupling and argue that $\lambda\phi_{4}^{4}$ theory
is a {}``marginal'' case between trivial $\lambda\phi_{d}^{4}$
theories with $d>4$ and theories with $d\le4$. The beta function
of the theory behaves as 
\be
\beta(\lambda)\sim\lambda
\label{eq:BetaLin}
\ee
at strong coupling $\lambda\gg1$ and, as a result, the renormalized coupling
$\lambda(\Lambda)$ diverges as $\Lambda\to\infty$ although the non-perturbative Landau pole is absent. 
Our framework also allows us to estimate scaling
violating corrections to the beta function. We find that they are
typically large in the regime where the asymptotics (\ref{eq:BetaLin})
is applicable. Finally, Section \ref{sec:Discussion} is devoted
to the discussion.

\section{De Sitter space as stochastic quantization machine\label{sec:stochastic}}

Let us consider a massive self-interacting scalar field theory with
Langrangian density
\be
{\cal L}=\frac{1}{2}(\partial\phi)^{2}-\frac{1}{2}m^{2}\phi^{2}-\frac{1}{4!}\lambda\phi^{4},
\label{eq:InitLagrangian}
\ee
in a $dS_{D}\times M_{d}$ spacetime. Here by $dS_{D}$ we understand
a $t\in[0,+\infty)$ part of the planar patch of the $D$-dimensional
de Sitter space covered by a coordinate system $(t,\mathbf{x})=(t,x_{1},\ldots,x_{D-1})$
with metric 
\be
ds^{2}=dt^{2}-a^{2}(t)\sum_{i=1}^{D-1}dx_{i}dx^{i}
\label{eq:dSmetric}
\ee
and scale factor $a(t)\sim\exp(Ht)$.%
\footnote{One can think of this geometry as the Hartle-Hawking one \cite{HartleHawking}
with a Euclidean sphere attached at the equator of de Sitter space corresponding
to $t=0$, though the difference from the Hartle-Hawking geometry
is that the patch (\ref{eq:dSmetric}) does not cover the $dS$ part
of the Hartle-Hawking geometry completely \cite{deSittercausal}.
The choice of this geometry automatically implies the choice of the
Bunch-Davies vacuum state for the quantum field theory (\ref{eq:InitLagrangian}).%
} The $d$-dimensional Euclidean space $M_{d}$ is either a torus $T_{d}$
or non-compact flat space $E_{d}$ covered by a coordinate system
$\mathbf{y}=(y_{1},\ldots,y_{d})$. The Hubble constant $H$ is related
to the curvature of $dS_{D}$ spacetime by $R=D(D-1)H^{2}$. In what
follows, we neglect the backreaction of the scalar field on background
spacetime, considering only the limit $M_{P}\to\infty$.

The dynamics of quantum Heisenberg operator $\phi(t,\mathbf{x},\mathbf{y})$
is determined by the equation of motion 
\be
\ddot{\phi}+(D-1)H\dot{\phi}-\frac{\nabla_{x}^{2}}{a^{2}}\phi-\nabla_{y}^{2}\phi+m^{2}\phi+
\frac{1}{6}\lambda\phi^{3}=0.
\label{eq:KGfreemassive1}
\ee
It is convenient to decompose the Heisenberg operator $\phi(t,\mathbf{x},\mathbf{y})$
into ultraviolet (subhorizon) and infrared (superhorizon) contributions
\cite{StarobinskyStochastic,StarobinskyStochastic2} according to
the prescription 
\be
\phi=\Phi+\int\frac{d^{d}p}{(2\pi)^{d/2}}\frac{d^{D-1}k}{(2\pi)^{\frac{D-1}{2}}}\theta(k-\epsilon aH)\left(a_{k,p}f_{k,p}(t)e^{-i\mathbf{(k}\mathbf{x}+\mathbf{py})}+a_{k,p}^{\dagger}f_{k,p}^{*}(t)
e^{i(\mathbf{kx}+\mathbf{py})}\right)+\delta\phi,
\label{eq:IR-UVdecomp}
\ee
where $\theta(x)$ is the Heaviside step function, $\epsilon$ is
a free parameter, 
\be
f_{k,p}(t)=\frac{\sqrt{-\pi\eta}}{2a^{\frac{D-2}{2}}}H_{\frac{D-1}{2}}^{(1)}(-k\eta)e^{\frac{i\pi D}{4}}
\label{eq:masslessmodes}
\ee
are modes of a free \emph{massless}\footnote{So that they do not depend on the momentum $p$ along extra dimensions.} scalar field in $dS_{D}$ background ($\eta=-\frac{1}{Ha(t)}$ is
conformal time) and $\delta\phi$ is a contribution of the order ${\cal O}\left({\rm max}\left(\frac{m^{2}}{H^{2}},\frac{\lambda\langle\phi^{2}\rangle}{H^{2}}\right)\right)$.
In order for the contribution $\delta\phi$ to be negligible, $\epsilon$
should satisfy the condition
\be
\exp\left(-\frac{(D-1)H^{2}}{m_{{\rm eff}}^{2}}\right)\ll\epsilon\lesssim1,
\ee
where $m_{{\rm eff}}^{2}=m^{2}+3\lambda(H)\langle\phi^{2}\rangle$
\cite{StarobinskyStochastic}. An estimation for the Hartree mass and
bounds for renormalized coupling will be given below.

In the leading order approximation in small {}``slow roll'' parameters
$\frac{m^{2}}{H^{2}},\,\frac{\lambda\langle\phi^{2}\rangle}{H^{2}}$,
the equation of motion for the infrared part $\Phi(t,\mathbf{x},\mathbf{y})$
of the Heisenberg operator $\phi(t,\mathbf{x},\mathbf{y})$ acquires
the form
\be
\dot{\Phi}=-\frac{1}{(D-1)H}(\nabla_{y}^{2}+m^{2})\Phi-\frac{1}{6(D-1)H}\lambda\Phi^{3}+f(t,\mathbf{x},\mathbf{y}),
\label{eq:Langevin}
\ee
where
\[
f(t,\mathbf{x},\mathbf{y})=\frac{\epsilon H^{2}}{(2\pi)^{(D+N-1)/2}a^{(D-4)/2}}\int d^{d}p\, d^{D-1}k\,\delta(k-\epsilon aH)\frac{i}{\pi}\Gamma\left(\frac{D-1}{2}\right)\times\]
\be
\times\left(-\frac{2}{k\eta}\right)^{\frac{D-1}{2}}\sqrt{-\frac{\pi\eta}{4}}\left(a_{k}e^{-ikx-ipy}e^{\frac{i\pi D}{4}}-a_{k}^{\dagger}e^{ikx+ipy}e^{-\frac{i\pi D}{4}}\right).
\label{eq:force}
\ee
The composite operator $f(t,\mathbf{x},\mathbf{y})$ has correlation
properties of white noise
\be
\langle f(t,\mathbf{x},\mathbf{y})f(t',\mathbf{x},\mathbf{y}')\rangle=
\frac{\Gamma\left(\frac{D-1}{2}\right)}{2\pi^{\frac{D+1}{2}}}H^{D-1}\delta(t-t')
\delta^{d}(\mathbf{y}-\mathbf{y}'),
\label{eq:RForceCorrProperties}
\ee
where average above is in the Bunch-Davies vacuum state. These correlation
properties do not depend on the parameter $\epsilon$ in the decomposition
(\ref{eq:IR-UVdecomp}) if the operators $f(t,\mathbf{x},\mathbf{y})$
and $f(t',\mathbf{x},\mathbf{y}')$ are taken at the same point $\mathbf{x}$
(or, more precisely, within the same Hubble patch) of the de Sitter
space $dS_{D}$. Another important property of equation (\ref{eq:Langevin})
is that all terms in it commutate with each other, and therefore the
large scale field $\Phi$ can be considered a classical quantity
\cite{StarobinskyStochastic}. Thus, the equation (\ref{eq:Langevin})
is a Langevin equation describing a random walk of the classical large
scale field $\Phi$.

Its physical meaning can be understood as follows. An observer living
in $dS_{D}\times M_{d}$ space cannot discriminate between different
modes with physical wavelengths $\lambda\sim\frac{a}{k}$ larger than
the scale of cosmological horizon $H^{-1}$ and interprets their collection
as a classical background field $\Phi(t)$. The value of $\Phi(t)$
changes, while more and more modes leave the cosmological horizon
(a mode $f_{k}$ with momentum $k$ leaves the horizon at the moment
of time $t$ such that $a(t)\sim\frac{k}{H}$).\footnote{An interesting example where modes do not become classical after horizon crossing is presented in \cite{Pascal}.} This picture is known
to describe the phenomenon of eternal (stochastic) inflation \cite{Eternalinf}.

The Langevin equation (\ref{eq:Langevin}) is analogous to the one
appearing in the procedure of stochastic quantization for a $D$-dimensional
quantum scalar field theory with potential $V(\phi)=\frac{1}{2}m^{2}\phi^{2}+\frac{1}{4!}\lambda\phi^{4}$,
implying that the late time behavior of the correlation functions of the
field $\phi$ propagating in $(D+d)$-dimensional $dS_{D}\times M_{d}$
spacetime is effectively determined by a $d$-dimensional theory. One
can say that for any interacting scalar field $\phi$, de Sitter space
{}``generates'' a Nicolai map \cite{Nicolai} from the field propagating
in the bulk of $dS_{D}$ to the same field at the boundary (horizon),
since the {}``random force'' operator $f(t,\mathbf{x},\mathbf{y})$
is a linear combination of modes $\delta\phi_{k}$ with momenta $k=aH$
(\ref{eq:force}).

We will now show how to construct the partition function of the corresponding
$d$-dimensional theory explicitly. The Fokker-Planck equation describing
evolution of the probability $P(\Phi(t,\mathbf{y}),t)$, which measures
a given value of the large scale field $\Phi$ in a given Hubble patch,
has the form
\be
\frac{\partial P}{\partial t}=\hat{H}P=\int d^{d}y\frac{\delta}{\delta\Phi}\left(A\frac{\delta}{\delta\Phi}+B(\nabla_{y}^{2}+m^{2})\Phi+\frac{\lambda B}{6}\Phi^{3}\right)P,
\label{eq:FokkerPlanck}
\ee
where
\be
A=\frac{H^{D-1}}{4\pi^{\frac{D+1}{2}}}\Gamma\left(\frac{D-1}{2}\right),\,\, B=\frac{1}{(D-1)H}.
\label{eq:AandB}
\ee
In order to illustrate several important properties of its general
solution, let us first perform a Fourier transform along the directions
in $M_{d}$ and neglect the self-interaction of the field $\Phi$, setting
$\lambda=0$ so that Kaluza-Klein modes with different momenta $p$
become decoupled from each other. The Fourier mode $P_{p}$ of the
probability distribution $P$ is given by 
\be
P_{p}(\Phi_{p},t)=\exp\left(-\frac{B}{4A}(p^{2}+m^{2})\Phi_{p}^{2}\right)\sum_{n=0}^{\infty}c_{n}\psi_{n}(\Phi_{p})
e^{-2AE_{n}(t-t_{0})},
\ee
where $\psi_{n}(\Phi_{p})$ and $E_{n}$ are eigenfunctions and eigenvalues
of the Schrodinger equation
\be
-\frac{1}{2}\psi_{n}''+\frac{(p^{2}+m^{2})^{2}B^{2}}{8A^{2}}\psi_{n}=
\tilde{E}_{n}\psi_{n}=\left(E_{n}+\frac{B(p^{2}+m^{2})}{4A}\right)\psi_{n}.
\ee
All eigenvalues $E_{n}$ are non-negative and the lowest eigenstate
has zero energy $E_{0}=0$, since the Hamiltonian in (\ref{eq:FokkerPlanck})
is supersymmetric \cite{StarobinskyStochastic}.%
\footnote{It is interesting to note that the full $dS_{D}$ space consisting
of two planar patches with $a(t)\sim\exp(\pm Ht)$ would actually
lead to \emph{two} Fokker-Planck hamiltonians $H_{+}$ and $H_{-}$,
with spectra unbounded from above and below correspondingly. This
is an issue well known in the method of stochastic quantization \cite{DamgaardHuffel}.%
} Therefore, all Fourier modes of the probability distribution $P$
have finite asymptotics
\be
P_{p}(\Phi_{p})=\exp\left(-\frac{B}{2A}(p^{2}+m^{2})\Phi_{p}^{2}\right)
\ee
at $t\to\infty$, e.g., in the vicinity of future infinity $I_{+}$
of de Sitter space. After performing an inverse Fourier transformation,
we finally find that the asymptotic probability to measure a given
value of the classical field $\Phi(\mathbf{y})$ in a given Hubble
patch and at a given point $\mathbf{y}$ in the extra dimensions is 

\begin{equation}
P(\Phi)=\prod_{p}P_{p}(\Phi_{p})=\exp\left(-\frac{C}{2}\int d^{d}y((\nabla_{y}\Phi)^{2}+m^{2}\Phi^{2})\right),
\end{equation}
where 
\begin{equation}
C=\frac{ 4\pi^{\frac{D+1}{2}} H^{-D} }{ (D-1)\Gamma\left(\frac{D-1}{2}\right) },
\label{eq:Cdefinition}
\end{equation}
or, introducing a new variable $\chi=\sqrt{C}\Phi$,
\be
P(\chi)=\exp\left(-\frac{1}{2}\int d^{d}y((\nabla_{y}\chi)^{2}+m^{2}\chi^{2})\right).
\label{eq:PartFunctionFree}
\ee
Similarly, one can show that, when self-interaction is taken into account,
\be
P(\chi)=\exp\left(-\frac{1}{2}\int d^{d}y((\nabla_{y}\chi)^{2}+m^{2}\chi^{2}+\frac{1}{12}\tilde{\lambda}\chi^{4})\right),
\label{eq:PartFunctionInt}
\ee
where
\be
\tilde{\lambda}=\frac{\lambda}{C}=\frac{(D-1)\Gamma\left(\frac{D-1}{2}\right)}{4\pi^{\frac{D+1}{2}}}\lambda H^{D}.
\label{eq:lambda4Ddef}
\ee
The correlation properties of the classical field $\chi(\mathbf{y})$
are calculated as in Euclidean field theory with partition function
\be
Z\{\chi\}=\int{\cal D}\chi\exp\left(-\frac{1}{2}\int d^{d}y((\nabla_{y}\chi)^{2}+m^{2}\chi^{2}+\frac{1}{12}\tilde{\lambda}\chi^{4})\right)
\label{eq:PartFunction}
\ee
We conclude that in the vicinity of the $I_{+}$ infinity, the interacting
scalar field theory in $dS_{D}\times M_{d}$ spacetime is reduced
to a Euclidean field theory of a field $\chi$ in $M_{d}$ manifold
in the sense that the probability to measure a given value of the
classical field $\Phi(\mathbf{y})$ in a given Hubble patch and at a 
point $\mathbf{y}$ in the extra dimensions is determined by the partition
function (\ref{eq:PartFunction}).%
\footnote{The constructed map is very different from the
dS/CFT correspondence \cite{dSCFT}, although it can also be considered a bulk/boundary correspondence. dS/CFT correspondence implies the existence
of non-local correlations in dS space, while our mapping implies ultralocality
of QFTs in the vicinity of future infinity $I_{+}$.%
}

Apart from the inverse correlation length $m_{{\rm ren}}$, there
are two scales of interest in the theory: $\frac{1}{2}\sqrt{D(D-2)}H\sim\frac{1}{2}DH$
and $\sqrt{2}H$. The inverse curvature radius of $dS_{D}$ space
$R_{dS}^{1/2}\sim DH$ plays the role of an ultraviolet cutoff for the
theory (\ref{eq:PartFunction}), since only the modes with sufficiently
low momenta $p$ along the extra dimensions can contribute to the partition
function (\ref{eq:PartFunction}). 

To prove this statement, let us consider the equation of motion for the modes $u_{k,p}=f_{k,p}a^{\frac{D-2}{2}}$
of a massive scalar field on a $dS_{D}\times M_{d}$ background
\be
u_{k,p}''+\left(k^{2}-\frac{D(D-2)-4(p^{2}+m^{2})/H^{2}}{4\eta^{2}}\right)u_{k,p}=0,
\label{eq:Umodes}
\ee
where prime denotes differentiation w.r.t. conformal time $\eta=-\frac{1}{Ha(t)}$.
The infrared effect we are after (build up of the coherent large scale
field $\Phi$) is due to the freeze-out of superhorizon modes with
$k\ll aH$, when the effective mass of the given mode $u_{k,p}$ becomes
negative. For a given $k$ and $p$, the solution of (\ref{eq:Umodes})
is a linear combination of a decaying and a growing mode. In the regime
$k\ll aH$, the decaying mode can be neglected compared to the growing
mode, so that the dynamics of $u_{k,p}$ becomes quasi-classical. Only
those modes contribute to the quasi-classical field $\Phi$, and 
the evolution of the latter is determined by the Langevin equation (\ref{eq:Langevin}).
Apparently, this quasi-classical regime can only be possible for
modes with momenta 
\be
p<\sqrt{p^{2}+m^{2}}\lesssim\frac{\sqrt{D(D-2)}}{2}H=L^{-1}.
\label{eq:CoarseGraining}
\ee
Therefore, the effective action in the partition function (\ref{eq:PartFunction})
is actually coarse-grained at scales $L\sim(DH)^{-1}$ determined
by (\ref{eq:CoarseGraining}). One can study the infrared and ultraviolet
properties of the theory (\ref{eq:PartFunction}) by appropriately modifying
this coarse-graining scale.

The inverse Hubble scale $H^{-1}$ is a natural unit for measuring
the time and length scales in the problem: it enters into the background
metric of spacetime only in the combination $\exp(Ht)$ and defines the
cross-correlation between different Hubble patches, since the equal
time pair correlation function of the field $\phi(t,\mathbf{x},\mathbf{y})$
has the form \cite{VilenkinCorrFunction} 
\be
\langle\phi(t,\mathbf{x},\mathbf{y})\phi(t,\mathbf{x}',\mathbf{y})\rangle\approx\langle\phi(t,\mathbf{x},\mathbf{y})
\rangle^{2}\left(1-\frac{1}{Ht}\log H|\mathbf{x}-\mathbf{x}'|\right).
\ee
Also, as we will see in the next Section, the scale $\sqrt{2}H$ provides
the fundamental ultraviolet cutoff for the theory in the sense that
at $p\lesssim\sqrt{2}H$, it can no longer be described by the partition
function (\ref{eq:PartFunction}).

\section{Physics at large momenta\label{sec:UVphysics}}

The partition function (\ref{eq:PartFunction}) correctly describes
physics of the $\lambda\phi^{4}$ theory in a $dS_{D}\times M_{d}$
spacetime at low physical momenta $\mathbf{p}_{{\rm phys}}^{2}=\frac{k^{2}}{a^{2}}+p^{2}\ll(DH)^{2}$
(and in the limit $a\to\infty$). The UV physics of the theory can be
analyzed straightforwardly by considering the behavior of the modes with
physical momenta $\mathbf{p}_{{\rm phys}}^{2}=\frac{k^{2}}{a^{2}}+p^{2}\gg(DH)^{2}$.
We are particularly interested in determining how this range of physical
momenta contributes to the renormalization of $\lambda$.

At $p\gg H$, $k\ll aH$, the behavior of the theory is the same as the
large scale behavior of \emph{heavy} scalar fields in a de Sitter background.
It is well known that such fields are not generated.%
\footnote{According to a recently presented alternative point of view \cite{Polyakov},
the vacuum of a self-interacting heavy scalar field theory in de Sitter
space (and the Hartle-Hawking geometry) and single particle states above
this vacuum are actually unstable. Our conclusions are different,
since we consider the planar patch of $dS$ space instead of the global
patch \cite{Polyakov}.%
} Indeed, in this regime one can rewrite the effective action for the field
$\phi$ in terms of the variable $v(t,\mathbf{x},\mathbf{y})=\phi(t,\mathbf{x},\mathbf{y})a^{\frac{D-1}{2}}$,
neglecting the $m^{2}\phi^{2}$ term and the dependence of the Lagrangian
density on the spatial coordinates $\mathbf{x}$ of $dS_{D}$ space:
\be
S\approx\int dtd^{D-1}xd^{d}y\left(\frac{1}{2}(\dot{v})^{2}-\frac{1}{2}(\nabla_{y}v)^{2}-\frac{\lambda v^{4}}{4!a^{D-1}}\right).
\ee
Therefore, the interval $p\gg H$, $k\ll aH$ only contributes to
the renormalization of $\lambda$ at early times $t\lesssim H^{-1}$.

Similarly, one can show that the interval $p\ll H$, $k\gg aH$ (corresponding to light
subhorizon modes) only contributes at $t\lesssim H^{-1}$: neglecting
the $m^{2}\phi^{2}$ term and the dependence on the coordinates $\mathbf{y}$
in $M_{d}$ space, one can rewrite effective action in terms of the
variable $u=\phi a^{\frac{D-2}{2}}$ as 
\be
S\approx\int d\eta d^{D-1}xd^{d}y\left(\frac{1}{2}(u')^{2}-\frac{1}{2}(\nabla_{x}u)^{2}-\frac{\lambda u^{4}}{4!a^{D-4}}\right)
\ee
and find that the interaction term is again suppressed by powers of the
scale factor. This suppression is present only for $D>4$, while at
$2\le D\le4$ we lose control over the behavior of the theory at $p\lesssim DH$.
This suggests that the scale $\sqrt{2}H$ is the fundamental UV cutoff
for the theory.

Finally, at $t\lesssim H^{-1}$ (and for the interval $p\gg H$, $k\gg aH$),
when all the modes are still within a single Hubble volume, the expansion
of the spacetime can be entirely neglected and the effective action
for the field $\phi$ acquires the form 
\be
S=\int dtd^{D-1}xd^{d}y\left(\frac{1}{2}(\dot{\phi})^{2}-\frac{1}{2}(\nabla_{x,y}\phi)^{2}-\frac{1}{2}m^{2}\phi^{2}
-\frac{1}{4!}\lambda\phi^{4}\right).
\label{eq:UVtheory}
\ee

The modes of the theory (\ref{eq:UVtheory}) propagate in flat
spacetime of dimensionality $D+d>4+d>D_{{\rm cr}}=4$ larger than
critical.%
\footnote{If the Hartle-Hawking geometry is considered, the action (\ref{eq:UVtheory})
should be replaced by the action of Euclidean theory in $S_{D}\times M_{d}$
space.%
} Therefore, all critical exponents of the theory can be determined
using mean field approximation \cite{ZunnJustinTextbook}. The $(D+d)$-dimensional
theory (\ref{eq:UVtheory}) is UV incomplete and should be provided
with a UV cutoff scale $\Lambda_{{\rm UV}}$. In the limit $\Lambda_{UV}\to\infty$
the theory (\ref{eq:UVtheory}) is trivial \cite{Aizenman1,Aizenman2},
so we are naturally interested in the case of finite $\Lambda_{{\rm UV}}$.

In the mean field approximation, the major effect of the interaction
term is the Hartree contribution to the effective mass of the field
$\phi$, which can be estimated as
\be
m_{{\rm eff}}^{2}=3\lambda_{0}\langle\phi^{2}\rangle\approx\frac{3\lambda_{0}\Lambda_{UV}^{D+d-2}}{4(D+d)},
\label{eq:HartreeMass}
\ee
where $\lambda_{0}$ is the bare coupling constant. If $\Lambda_{{\rm UV}}$
is sufficiently large, $\lambda\langle\phi^{2}\rangle\gg m^{2}$.
In order for the infrared limit (\ref{eq:PartFunction}) to exist,
the condition $m_{{\rm eff}}^{2}\lesssim L^{-2}=\frac{D(D-2)}{4}H^{2}$
or 
\be
\lambda_{0}\Lambda_{{\rm UV}}^{D+d-4}\lesssim\frac{4(D+d)}{3}\frac{1}{(\Lambda_{{\rm UV}}L)^{2}},
\label{eq:HartreeBound1}
\ee
should be satisfied. The physical meaning of this condition is simple:
in the theory (\ref{eq:PartFunction}) with UV cutoff $L^{-1}$, modes
with effective mass larger than the cutoff scale are integrated out.

Non-gaussian contributions to correlation functions of the theory
(\ref{eq:UVtheory}) are expected to be small in the limit of large
$\Lambda_{{\rm UV}}$, since the renormalized 4-point vertex (and
the renormalized self-interaction constant $\lambda$) remains uniformly
bounded (and vanishes in the limit $\Lambda_{{\rm UV}}\to\infty$)
\cite{Aizenman1,Aizenman2} as 
\be
\lambda_{{\rm ren}}=\frac{|u^{(4)}(p_{i}=0)|}{G^{4}(p=0)}\le\frac{12(D+d)}{\Lambda_{{\rm UV}}^{D+d-4}}\left(1-e^{-C_{D+d}\lambda_{0}\Lambda_{UV}^{D+d-4}}\right)e^{\frac{2m_{{\rm eff}}}{\Lambda_{{\rm UV}}}}\sim\Lambda_{{\rm UV}}^{-(D+d-4)},
\label{eq:Aizenman}
\ee
where $u^{(4)}(p_{i}=0)$ is the irreducible four-point function calculated
at zero external momenta, $G(p=0)$ is the two-point function at zero
momentum and $C_{D+d}=\frac{1}{2}\left(\int_{-\pi}^{\pi}\frac{d^{D+d}p}{(2\pi)^{D+d}}\left(\sum_{i=1}^{D+d}4\sin^{2}(p_{i}/2)\right)^{-1}\right)^{2}<\infty$.

This justifies the approximations used to derive the Langevin equation
(\ref{eq:Langevin}) in Section \ref{sec:stochastic}. The decomposition
(\ref{eq:IR-UVdecomp}) exists because the effective masses of modes are
required to be small compared to the coarse-graining scale (\ref{eq:CoarseGraining}).
Also, since the renormalized coupling $\lambda_{{\rm ren}}$ calculated
at the coarse-graining scale (\ref{eq:CoarseGraining}) is small,
one can express the composite operator $f(t,\mathbf{x},\mathbf{y})$
as a linear superposition of modes $f_{k,p}(\eta)$.

\section{Critical behavior at low and high momenta\label{sec:Critical}}

We are finally ready to discuss the critical behavior of the theory
(\ref{eq:PartFunction}). From now on, we will mostly focus on the
case $d=4$, when the theory (\ref{eq:PartFunction}) is renormalizable.

In the infrared region, where the theory is effectively reduced to
(\ref{eq:PartFunction}), the beta-function and the renormalized coupling
constant $\tilde{\lambda}(L^{-1})$ satisfy the equations (in
1-loop approximation)
\be
\beta(\tilde{\lambda})=\frac{d\tilde{\lambda}}{d\log L^{-1}}=\frac{3}{16\pi^{2}}\tilde{\lambda}^{2},
\label{eq:1loopBeta}
\ee
\be
\tilde{\lambda}'(\Lambda')=\frac{\tilde{\lambda}(\Lambda)}{1-\frac{3}{16\pi^{2}}\tilde{\lambda}(\Lambda)
\log(\frac{\Lambda'}{\Lambda})},
\label{eq:RenormalizationGroupWeak}
\ee
where $\Lambda=1/L$, $\Lambda'=1/L'$ and $L^{-1}=\frac{\sqrt{D(D-2)}}{2}H$
is the coarse-graining scale. This renormalization law can be straightforwardly
derived by either analyzing the infrared partition function (\ref{eq:PartFunction})
or performing a diagrammatic expansion of the solution of the Langevin
equation (\ref{eq:Langevin}) (see \cite{NamikiStochastic,StochasticQRenorm}
for the details).

When $\Lambda$ approaches the fundamental cutoff $\sqrt{2}H$, one
should expect strong deviations from (\ref{eq:RenormalizationGroupWeak}),
since modes along the $D$ dimensions of de Sitter space become excited
and the theory can no longer be described by the partition function
(\ref{eq:PartFunction}). In our setup, the behavior of the beta-function
$\beta=\frac{\partial\tilde{\lambda}}{\partial\log\Lambda}$ (and
the renormalized coupling $\tilde{\lambda}$) at large momenta $\Lambda\gtrsim\sqrt{2}H$
can be found using the fact that at $\Lambda\sim\sqrt{2}H$, the correlation
functions of the 4-dimensional theory with partition function (\ref{eq:PartFunction})
are by construction smoothly transformed into correlation functions
of the $(D+4)$-dimensional theory (\ref{eq:UVtheory}), which can be
calculated using mean field approximation (MFA).

One way to estimate the {}``renormalized coupling'' $\lambda_{{\rm ren}}$
below the coarse-graining scale is to use the Aizenman's inequality
(\ref{eq:Aizenman}) at the saturation point. Since according to (\ref{eq:HartreeBound1})
one has $m_{{\rm eff}}\ll L^{-1}\ll\Lambda_{{\rm UV}}$, the exponent
$\exp(2m_{{\rm eff}}/\Lambda_{{\rm UV}})$ in the r.h.s. of (\ref{eq:Aizenman})
can be omitted. Also, only values of $\lambda_{0}$ such that $\lambda_{0}\Lambda_{{\rm UV}}^{D}\gg1$
are of physical relevance: larger $\lambda_{0}$ correspond to larger
renormalized coupling $\tilde{\lambda}(L^{-1})$ of the theory (\ref{eq:PartFunction}),
and we are interested to study the correlation properties of the latter
at $\tilde{\lambda}>1$.%
\footnote{The only bound on possible values of $\lambda_{0}$ is the Hartree
bound (\ref{eq:HartreeBound1}). Apparently, for any $\Lambda_{{\rm UV}}$
it is possible to choose $\lambda_{0}$ such that $\lambda_{0}\Lambda_{{\rm UV}}^{D}\gg1$
by appropriately tuning the values of $H$ and $D$.%
} Therefore, at $\lambda_{0}\Lambda_{UV}^{D}\gg1$ one has a bound
on $\lambda_{{\rm ren}}$ independent of $\lambda_{0}$:
\be
\lambda_{{\rm ren}}\lesssim\frac{12(D+4)^{2}}{\Lambda_{{\rm UV}}^{D}}.
\ee
Rewriting it in terms of the effective 4-dimensional coupling (\ref{eq:lambda4Ddef}),
we have
\be
\tilde{\lambda}_{{\rm ren}}\lesssim\frac{3(D+4)^{2}(D-1)\Gamma\left(\frac{D-1}{2}\right)2^{D}}{\pi^{\frac{D+1}{2}}(D(D-2))^{D/2}}(\Lambda_{{\rm UV}}L)^{-D}.\label{eq:AizenmanMaximal}
\ee

To derive a similar bound for the case of small $\lambda_{0}$, it
is convenient to present the Aizenman's inequality (\ref{eq:Aizenman})
in a somewhat different form \cite{Aizenman2}:
\be
\lambda_{{\rm ren}}\le\frac{3\Lambda_{{\rm UV}}^{D+4}}{\langle\phi^{2}\rangle_{0}^{2}}\left(1-e^{-\hat{\lambda}_{0}\langle\phi^{2}\rangle_{0}^{2}}\right),
\label{eq:AizenmanMod}
\ee
where
\be
\langle\phi^{n}\rangle_{0}=\int_{-\infty}^{+\infty}d\phi\phi^{n}\exp\left(-B\phi^{2}-\frac{1}{4!}\hat{\lambda}_{0}\phi^{4}\right)
/\int_{-\infty}^{+\infty}d\phi\exp\left(-B\phi^{2}-\frac{1}{4!}\hat{\lambda}_{0}\phi^{4}\right),
\label{eq:MFA}
\ee
\be
B\approx\frac{2(D+4)}{\Lambda_{{\rm UV}}^{D+2}},\,\,\,\,\hat{\lambda}_{0}=\frac{\lambda_{0}}{\Lambda_{{\rm UV}}^{D+4}}.
\ee
Since we require the condition (\ref{eq:HartreeBound1}) to hold,
the dimensionless parameter $\hat{\lambda}_{0}/B^{2}$ is small for
almost any value of $D$, and the integral (\ref{eq:MFA}) can be
calculated in the Gaussian approximation revealing 
\be
\tilde{\lambda}_{{\rm ren}}\lesssim3\lambda_{0}\left(1-\frac{\lambda_{0}\Lambda_{{\rm UV}}^{D}}{4(D+4)^{2}}\right)\frac{(D-1)\Gamma\left(\frac{D-1}{2}\right)2^{D+2}}{\pi^{\frac{D+1}{2}}(D(D-2))^{D/2}}
(\Lambda_{{\rm UV}}L)^{-D}.
\label{eq:AizenmanWeak}
\ee

Another way to determine $\tilde{\lambda}_{{\rm ren}}(L^{-1}>\sqrt{2}H)$
is to use MFA and the standard definition of the renormalized coupling,
relating it to 4- and 2-point irreducible correlation functions
\be
\lambda_{{\rm ren}}
=\frac{(3\langle\phi^{2}\rangle^{2}-\langle\phi^{4}\rangle)\Lambda_{{\rm UV}}^{D+4}}{\langle\phi^{2}\rangle^{4}}.
\label{eq:RenMFA}
\ee
Then, a straightforward calculation shows that 
\begin{equation}
\tilde{\lambda}_{{\rm ren}}
\approx \tilde{c}_{1} \frac{(D-1)\Gamma\left(\frac{D-1}{2}\right)2^{D}}{\pi^{D/2}(D(D-2))^{D/2}}
\lambda_{0}L^{-D}-\tilde{c}_{2}(D)\lambda_{0}^{2}\Lambda_{{\rm UV}}^{D}L^{-D}+{\cal O}
(\lambda_{0}^{3}\Lambda_{{\rm UV}}^{2D}L^{-D}).
\label{eq:MFArencoupling}
\end{equation}
The functional dependence of the renormalized coupling $\tilde{\lambda}$
on the scale $L^{-1}$, a direct consequence of (\ref{eq:lambda4Ddef}),
is of course similar for all three estimates (\ref{eq:AizenmanMaximal}),
(\ref{eq:AizenmanWeak}), (\ref{eq:MFArencoupling}): 
\begin{equation}
\tilde{\lambda}(L^{-1})\sim\frac{(D-1)\Gamma\left(\frac{D-1}{2}\right)2^{D}}{\pi^{D/2}(D(D-2))^{D/2}}(\Lambda_{{\rm UV}}L)^{-D}.
\label{eq:LambdaFunctionL}
\end{equation}
Note that we have to keep the dependence on $D$ in this expression
since $D$ itself is a function of the fundamental cutoff $\sqrt{2}H$
and the coarse-graining scale $L^{-1}$:
\be
D=1+\sqrt{1+\frac{4\Lambda^{2}}{H^{2}}}.
\label{eq:Ddef}
\ee
Differentiating (\ref{eq:LambdaFunctionL}) w.r.t. the coarse-graining
scale $\Lambda=L^{-1}$, we finally find the following expression for
the beta-function at large momenta:
\be
\beta(\tilde{\lambda})=\frac{d\tilde{\lambda}}{d\log\Lambda}\approx\tilde{\lambda}\left(\frac{D(D-2)}{2(D-1)}\left(\frac{1}{D-1}+\frac{1}{2}\psi\left(\frac{D-1}{2}\right)+\log\frac{H}{\sqrt{\pi}\Lambda_{{\rm UV}}}\right)+D\right),
\label{eq:BetaFunc}
\ee
where $\psi(x)=\frac{\Gamma'(x)}{\Gamma(x)}$.

\section{Discussion\label{sec:Discussion}}

Naturally, we are interested to find renormalization group trajectories
such that
\begin{enumerate}
\item the fundamental cutoff of the theory (\ref{eq:PartFunction}) is well
below the Landau pole $\Lambda=\Lambda_{{\rm probe}}\exp\left(\frac{3}{16\pi^{2}\tilde{\lambda}^{2}(\Lambda_{{\rm probe}})}\right)$, where $\Lambda_{{\rm probe}}$ is some {}``probe'' momentum scale
and $\tilde{\lambda}(\Lambda_{{\rm probe}})$ is the value of renormalized
coupling which an observer measures testing physics at scales $\sim\Lambda_{{\rm probe}}$,
\item the coupling $\tilde{\lambda}(\Lambda\approx\sqrt{2}H)$ is large (if
possible, even larger or comparable to the tree level unitarity bound
$\tilde{\lambda}=\frac{32\pi}{\sqrt{3}}$), and
\item scaling violations \cite{Symanzik} are still small at $\Lambda\sim\sqrt{2}H$.
\end{enumerate}
The existence of such RG trajectories within our framework would provide
a {}``proof-of-concept'' that a strong coupling regime for the $\lambda\phi_{4}^{4}$
theory does exist. Let us see whether such RG trajectories are present
in our construction.

Naively, one can expect the transition between the regimes (\ref{eq:1loopBeta}),
(\ref{eq:RenormalizationGroupWeak}) and (\ref{eq:BetaFunc}), (\ref{eq:MFArencoupling})
to happen around $\Lambda\approx\sqrt{2}H$ (corresponding to $D\approx4$),
the point coinciding with the fundamental UV cutoff for the theory
(\ref{eq:PartFunction}) in our construction (see the discussion in the
Section \ref{sec:UVphysics}). As follows from the expression (\ref{eq:BetaFunc}),
scaling violations are already significant at this point (see (\ref{eq:Ddef})):
\be
\beta(\tilde{\lambda})\approx3.7\tilde{\lambda}+\frac{4}{3}\log\frac{H}{\Lambda_{{\rm UV}}},
\label{eq:Beta4}
\ee
so the beta-function has non-universal contributions (although the first
term in the r.h.s. of (\ref{eq:Beta4}) is clearly universal). This
result seems to confirm the conclusion of \cite{TrivialityArgumentsRG}.

On the other hand, one can argue that the matching can actually happen
at scales much lower than the fundamental cutoff $\sqrt{2}H$. Indeed,
although we lose control over our construction at $\Lambda\ll\sqrt{2}H$
(or, equivalently, $D\to2$), it is still possible to take the {}``infrared''
limit $\Lambda\ll\sqrt{2}H$ of the expression (\ref{eq:BetaFunc}),
which behaves sufficiently well as a function of parameter $D$. In
the leading approximation one finds 
\be
\beta(\tilde{\lambda})\approx2\tilde{\lambda},
\label{eq:BetaScaling}
\ee
while scaling violating terms are suppressed by powers of the
small parameter $\Lambda^{2}/H^{2}$. This is exactly the functional
dependence on the RG energy scale and cutoff that such terms are expected
to have \cite{Symanzik}!

The expression (\ref{eq:BetaScaling}) for the beta-function seems
to be rather surprising. First of all, it implies the absence of a Landau
pole: indeed, from (\ref{eq:BetaScaling}) one has
\be
\tilde{\lambda}\sim\Lambda^{2},
\ee
so that there is no blow up at finite momentum, although the renormalized
coupling grows without bound at $\Lambda\to\infty$. Second, we note
that the linear behavior $\beta(\tilde{\lambda})\sim\tilde{\lambda}$
of the beta function was found in \cite{Suslov} by resummation of
Lipatov instantons. As we know, such resummation leads to a correct
result only in the absence of renormalons \cite{tHooft}, so that
linear behavior of the beta function at large coupling suggests the
absence of renormalons in $\lambda\phi_{4}^{4}$ theory \cite{Renormalons}.

\begin{figure}[!ht]
\begin{center}
\includegraphics{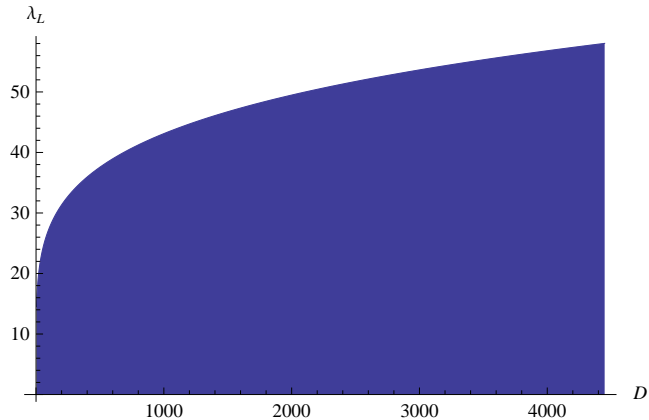}
\caption{The value of the renormalized coupling $\tilde{\lambda}$ at the matching point as a function of the parameter $D$. The shaded area corresponds to the weak coupling regime of the theory.\label{fig1}}
\end{center}
\end{figure}

The value of the renormalized coupling $\lambda_{*}$ at the matching
point between the regimes (\ref{eq:1loopBeta}), (\ref{eq:RenormalizationGroupWeak})
and (\ref{eq:BetaFunc}), (\ref{eq:LambdaFunctionL}) can be found
from the matching condition for the beta function
\be
D\lambda_{*}\approx\frac{3}{16\pi^{2}}\lambda_{*}^{2}-\frac{17}{768\pi^{4}}\lambda_{*}^{4}+\ldots,
\label{eq:MatchingBetaFunction}
\ee
where the l.h.s. of this equality represents the universal contribution to the beta
function (\ref{eq:BetaFunc}). After the substitution of the expression
for the beta function up to 3 loops \cite{4loop} into the r.h.s.
of (\ref{eq:MatchingBetaFunction}), we find $\lambda_{*}(D=2)\approx12.66$
and $\lambda_{*}(D=4)\approx14.53$. The first value corresponds to
the situation when matching takes place at small momenta $\Lambda\ll\sqrt{2}H$, while
the second to matching at $\Lambda\approx\sqrt{2}H$. Both values
$\lambda_{*}$ are well below the tree level unitarity bound but above
the region of validity of the 1-loop approximation for the beta function.
Generally, the function $\lambda_{*}=\lambda_{*}(D)$ is a slowly
changing function of the parameter $D$, finally reaching tree level
unitarity bound around $D\approx{}4450$, see Fig. \ref{fig1}. Since the parameter
$D$ determines the order of magnitude of the scaling violation corrections,
we see that the strong coupling regime with $\lambda_{*}$ of the order
of the tree level unitarity bound is realized when scaling violations
are already very strong, confirming the main conclusion of \cite{TrivialityArgumentsRG}:
a truly strongly coupled regime with $\lambda\ge\frac{32\pi}{\sqrt{3}}$
and negligible scaling violation cannot be achieved in $\lambda\phi^4_4$ theory.

Finally, we would like to point out that the formalism developed in
Sections \ref{sec:stochastic}, \ref{sec:UVphysics} and \ref{sec:Critical}
is perfectly applicable for the analysis of the high energy behavior of
the $\lambda\phi^{4}$ theory in $d=2,3$ dimensions, where it leads to the scaling behavior 
\[
\beta(\tilde{\lambda})\approx d\tilde{\lambda}
\]
in the transient regime. For $d=2$ the value of the beta function
coincides with the exact result \cite{Ising}.

\bigskip
\bigskip

\noindent
{\bf \large Acknowledgments}

\bigskip

I would like to thank P.H. Damgaard, L. Motl and A. Polyakov for comments.
E. Greenwood, T. Vachaspati and P. Vaudrevange have made numerous
valuable suggestions thats helped to improve the text of the paper.
This work is supported by grants from NASA and US DOE to the particle
astrophysics theory group at CWRU.

\end{document}